\documentclass[12pt]{article}
\usepackage{graphicx}
\begin{document}
\title{\textbf{DIRECT QUANTIZATION OF OPEN DISSIPATIVE SYSTEMS}}
\author{Jose A. Magpantay \thanks{email: jamag@nip.upd.edu.ph}\\National Institute of Physics\\University of the Philippines, Diliman Quezon City, 1101 Philippines}
\maketitle
\begin{abstract}
Open, dissipative systems subject to a random force are directly quantized.  The starting point is the effective action derived using the method of Parisi-Sourlas.  Since the effective action is second-order, the method of Ostrogradsky was used to quantize the system canonically.  In the case of the harmonic oscillator, the relevant Green function can be computed exactly.  In the general case, a perturbation expansion, involving time-dependent (memory) terms, can be defined.
\end{abstract}
\clearpage
	In this paper we will show how to directly quantize open, dissipative systems that are subject to a random force.  Previous efforts in this direction were undertaken more than twenty years ago by Kostin \cite{kostin} and Yasue \cite{yasue}.  Kostin made use of a non-linear Schroedinger equation, which was later derived using Nelson's quantization \cite{nelson} method by Yasue.  Since then open systems have been quantized by ``closing'' the system first, which means assuming a particular model of the environment and its interaction with the system \cite{cald}.  Given the Hamiltonian for the system and environment, quantization is straightforward.  By integrating out the degrees of freedom of the environment, the quantum theory of the system is derived.

	The reason why a direct quantization of an open, dissipative system subject to a random force is not possible is, supposedly, the absence of a Langrangian or a Hamiltonian.  However, as we will show in the next section,  there is an effective Lagrangian appropriate to such a system.  This Lagrangian can be derived using the method of Parisi and Sourlas \cite{parisi}.  Starting with this Lagrangian, we quantize the system using the canonical path-integral framework.

	The equation of motion for an open, dissipative system subject to random force is
\begin{equation}\label{open}
m \frac{d^2 x}{dt^2} + \eta \frac{dx}{dt} + \frac{dU}{dx}=f(t)
\end{equation}
where $m$, $\eta$ and $U(x)$ are, respectively, the mass, friction constant and potential of the particle.  Because of the friction term, the LHS of equation (\ref{open}) is not derivable from an action, i.e., there is no $S_o$ such that
\begin{equation}\label{friction}
\frac{\delta S_o}{\delta x}=m \frac{d^2 x}{dt^2} + \eta \frac{dx}{dt} + \frac{dU}{dx}.
\end{equation}
For this reason, a direct quantization of the system using the canonical/path-integral method supposedly is not possible.

	But is there really no action that describes the system?  Note that although there is no solution to equation (\ref{friction}), the system is not described by (2) but by equation (\ref{open}) plus the condition 
\begin{equation}\label{condition}
\left<f(t)f(t')\right>=i\delta(t-t').
\end{equation}
Equation (\ref{condition}) implies that the distribution of $f$  is gaussian, i.e.,
\begin{equation}\label{gaussian}
P[f]\sim e^{\frac{i}{2} \int dt\, f^2 (t)}.
\end{equation}
Following Parisi and Sourlas' arguments \cite{parisi}, we find 
\begin{equation}\label{random}
\left<x_f(t)x_f(t')\right>_f= \int (dx) x(t) x(t') e^{\frac{i}{2} \int dt\left( \frac{\delta S_o}{\delta x}\right)^2 + tr \ell{n} \left( \frac{\delta^2 S_o}{\delta x^2}\right) }.
\end{equation}

At the LHS, $x_f(t)$  refers to the solution of equation (\ref{open}) and the average over $f$ makes use of the distribution given by equation (\ref{gaussian}).  The RHS of equation (\ref{random}) therefore is a classical two-point function of a particle in a random field.  From this we read the effective action of an open, dissipative system subject to a random force as
\begin{equation}\label{quantize}
S_o(x)=\frac{1}{2}\int dt\left(m \frac{d^2x}{dt^2}+\eta \frac{dx}{dt}+\frac{dU}{dx}\right)^2 - itr \ell{n} \left( \frac{\delta^2 S_o}{\delta x^2}\right). 
\end{equation}
This will be the starting point of  the quantization process.

Let us quantize a specific system.  Take the case of the harmonic oscillator, with potential $U(x) =\frac{1}{2}kx^2$ .  We can neglect the second term in equation (\ref{quantize}) because $\frac{\delta^2 S_o}{\delta x^2}$   is $x$ independent.  The resulting action is second-order and time-independent.  The method of Ostrogradsky \cite{gross} is clearly applicable in this case.

	Define the coordinates and momenta
\begin{eqnarray}
q_1&=&x, \label{action}\\
q_2&=&\dot{x}, \label{coord}\\
p_1&=&\frac{\partial L}{\partial \dot{x}}- \frac{d}{dt}\left(\frac{\partial L}{\partial \ddot{x}}\right)= \eta[m \ddot{x}+ \eta \dot{x}+kx]-m[m \ddot{q_2}+\eta \ddot{x}+k \dot{x}], \label{momenta}\\
p_2&=&\frac{\partial L}{\partial \ddot{x}}=m \left[m\ddot{x}+ \eta \dot{x}+ kx \right]. \label{motion}
\end{eqnarray}
The Hamiltonian is given by
\begin{equation}\label{classical}
H = \sum_{i=1}^2 p_i\dot{q}_i-L=\frac{1}{2}\frac{p_2^2}{m^2}-\frac{\eta}{m}p_2q_2-\frac{k}{m}p_2q_1+p_1q_2 .
\end{equation}
Note that the classical equations of motion in the Hamiltonian formalism given by
\begin{eqnarray}
\dot{q}_i&=&\frac{\partial H}{\partial p_i}, \label{Hamil}\\
\dot{p}_i&=&- \frac{\partial H}{\partial q_i}, \label{form}
\end{eqnarray}
give the Euler-Lagrange equation,
\begin{equation}\label{velocity}
\frac{d^2}{dt^2}\left(\frac{\partial L}{\partial \ddot{x}}\right)-\frac{d}{dt}\left(\frac{\partial L}{\partial \dot{x}}\right)+\frac{\partial L}{\partial x}= 0.
\end{equation}

Since one of the coordinates  $(q_2)$  is a velocity, it is a good idea to check the consistency of the Poisson brackets before proceeding with the quantization.  The fundamental brackets are
\begin{eqnarray}
\left[q_i,q_j\right]&=&\left[p_i,p_j\right]=0, \,\,\,\,\,\, {i,j = 1,2,} \label{jacob}\\
\left[q_i,p_j\right]&=&\delta_{ij} \,\,\,\,\,\,\,\,\,\,\,\,\,\,\,\,\,\,\,\,\,\,\,\,\,\,\,\,\,{i,j = 1, 2}.\label{poison}
\end{eqnarray}
These brackets are consistent as shown by the fact that the equation of motion follows from them.  Also the Jacobi identity
\begin{eqnarray*}
\lbrack {A,\lbrack {B,C\rbrack}\rbrack}+\lbrack {B,\lbrack {C,A\rbrack}\rbrack}+\lbrack {C,\lbrack {A,B\rbrack}\rbrack}=0,
\end{eqnarray*}
for  any $A$, $B$ and $C$ function of $q_i$ and $p_j$ follows  from the Poisson brackets. 

	We can now quantize the system  using the canonical/path-integral method.  This results in the Hamiltonian path-integral given by
\begin{eqnarray*}
\int(dq_1)(dq_2)(dp_1)(dp_2)e^{\frac{i}{\hbar}\int dt 
\lbrack{p_1}\dot{q_1}+{p_2}\dot{q_2}-H{(q_i,p_i)}\rbrack}.
\end{eqnarray*}
The only minor complication in the derivation of the above expression is the operator-ordering of the $p_2q_2$ term, which was resolved by symmetrization.  Integrating out all terms except $q_1 = x$ yields
\begin{equation}\label{integral}
PI = \int(dx)e^{\frac{i}{\hbar}\int dt\frac{1}{2}(m\ddot{x}+\eta \dot{x}+kx)^2}.
\end{equation}
But this is just the path-integral given in equation (\ref{random}), the only difference being the appearance of  $\hbar$  in  equation (\ref{integral}).  Thus, the stochastic dynamics of an open dissipative system is similar to its quantum dynamics.  That this is true is supported by the work of Haba \cite{zaba}, who showed that quantum open systems are random classical dynamical systems.

	Since the effective action is quadratic, the two-point function is exact and given by the Green function of the operator 
$m^2\frac{d^4}{dt^2}+\left(2mk-\eta^2\right)\frac{d^2}{dt^2}+k^2$, which is
\begin{equation}\label{operator}
g(t-t')=\frac{1}{2\pi}\int_{- \infty}^{\infty}d\omega \frac{e^{i \omega (t-t')}}{m^2\omega^4-\omega^2(2mk-\eta^2)+k^2}
\end{equation}
The contour evaluation of equation (\ref{operator}) shows six distinct regimes:
\begin{eqnarray*}
(i) & & m>k+ \frac{\eta^2}{2k},\label{contour}\\
(ii) & & m=k+\frac{\eta^2}{2k},\label{regime}\\
(iii) & & \frac{\eta^2}{2k}<m<\frac{\eta^2}{2k}+k,\label{green}\\
(iv) & & \frac{\eta^2}{2k} - k<m< \frac{\eta^2}{2k},\label{quadra}\\
(v) & & m=\frac{\eta^2}{2k}-k,\label{omega}\\
(vi) & & m<\frac{\eta^2}{2k}-k.
\end{eqnarray*}
For example, in the case of $(iii)$, the Green function is
\begin{equation}\label{poten}
g(t-t') = \frac{m}{k^3}\left(\frac{1}{4}{\alpha^2\beta^2}\right) e^{-\frac{k}{m}\beta(t-t')}\sin \left[{\alpha(t-t')+\tan^-1}\left({\frac{\alpha}{\beta}}\right)\right],
\end{equation}
where
\begin{eqnarray}
\alpha=\frac{1}{2}+ {\frac{2mk-\eta^2}{4k^2}},\label{depend}\\
\beta=\frac{1}{2}-{\frac{2mk-\eta^2}{4k^2}}. \label{time}
\end{eqnarray}
	
	Next we consider potentials of the form
\begin{equation}\label{delta}
U(x) = \frac{1}{2}kx^2 + V(x),
\end{equation}
where $V(x)$ is $0(x^3)$ or higher.  In this case
\begin{equation}\label{case}
\frac{\delta^2 S_o}{\delta x^2}=m\frac{d^2}{dt^2}+\eta\frac{d}{dt}+k+\frac{d^2V}{dx^2}.
\end{equation}
The second term of equation (\ref{quantize}) is time-dependent and represents memory of the particle's interaction with the environment as shown by the expansion
\begin{equation}\label{envi}
tr\ell{n}\frac{\delta^2S_o}{\delta x^2}\approx\int dt h(0)\frac{d^2 V}{dx^2}(x(t))-\frac{1}{2}\int dt dt'h(t-t')\frac{d^2V}{dx^2}(x(t'))h(t'-t)\frac{d^2V}{dx^2}(x(t))+ \dots
\end{equation}
In equation (\ref{envi}), $h(t-t')$  is the Green function of the operator $m\frac{d^2}{dt^2}+\eta\frac{d}{dt}+k$ and is given by
\begin{equation}\label{integ}
h(t-t')=\frac{1}{2\pi}\int^{\infty}_{-\infty}d\omega\frac {e^{i\omega(t-t')}}{-m\omega^2+i\eta\omega+k}.
\end{equation}
The contour integration of equation (\ref{integ}) yields
\begin{equation}\label{linear}
h(t-t')= \left\{\begin{array}
	{r@{\quad\quad}l}
	\frac{e^{-\frac{\eta}{m}(t-t')}}{\left(4km-\eta^2\right)^{\frac{1}{2}}} \sin \left[\frac{1}{m}\left(4km-\eta^2\right)^{\frac{1}{2}}(t-t')\right],& \mbox{for}\,\,4km>\eta^2,\\ & \\

	 \frac{1}{2\pi m}\left(t-t'\right)e^{-\frac{\eta}{m}(t-t')},&  \mbox{for}\,4km = \eta^2,\\ & \\

	 \frac{e^{-\frac{\eta}{m}(t-t')}}{\left(-4km+\eta^2\right)^{\frac{1}{2}}} \sinh \left[\frac{1}{m}\left(-4km+\eta^2\right)^{\frac{1}{2}}(t-t')\right],&  \mbox{for}\,4km < \eta^2.
	\end{array}\right.
\end{equation}
The propagators correspond  to the underdamped, critically damped and overdamped cases. 

	The derivation of the effective action shows that in the non-linear case, the system's memory of its interaction with the environment is given by explicit time-dependent terms.  But in the case of the harmonic oscillator, the cumulative effect of the system's interaction with the environment is ``hidden'' in a propagator which is a Green function of a fourth-order operator  $m^2\frac{d^4}{dt^4} + \left(2mk - \eta^2\right) \frac{d^2}{dt^2} + k^2$  instead of the expected second-order operator $m\frac{d^2}{dt^2}+\eta\frac{d}{dt}+k$.

	To go around the problem of dealing with an explicitly time-dependent system, we use fermionic degrees of freedom to express the determinant in equation (\ref{random}).  The effective action we will deal with is
\begin{eqnarray}\label{freedom}
S[x,\phi,\bar\phi]&=&\frac{1}{2}\int dt\left\{\left(m\frac{d^2x}{dt^2}+\eta\frac{dx}{dt}+kx+\frac{dV}{dx}\right)^2+m\frac{d\bar\phi}{dt}\frac{d\phi}{dt}\right.\nonumber\\
& &+\left.\eta\bar\phi\frac{d\phi}{dt}+\bar\phi\left(k+\frac{d^2V}{dx^2}\right)\phi\right\}.
\end{eqnarray}
Again following the method of Ostrogradsky, the momenta are those given by equations (\ref{momenta}),(\ref{motion}) and
\begin{eqnarray}
\pi_\phi &=& -m\frac{d \bar\phi}{dt} - \eta\bar\phi, \label{ostro}\\
\pi_{\bar \phi} &=& m\frac{d\phi}{dt}. \label{method}
\end{eqnarray}
The Hamiltonian is
\begin{equation}\label{integrate}
H' = H +\frac{1}{m}\pi_\phi\pi_{\bar\phi}-\frac{1}{m}\pi_{\bar\phi}(\pi_\phi+\eta\bar\phi)+\frac{1}{m}(\pi_\phi+\eta\bar\phi)\pi_{\bar\phi}+\frac{1}{m}\eta\bar\phi\pi_{\bar\phi}+{\bar\phi}\left(k+ \frac{d^2V}{dx^2}\right)\phi .
\end{equation}
From this Hamiltonian, we construct the Hamiltonian path-integral, integrate the momenta and the coordinates $\phi$  and $\bar\phi$ to arrive at the path-integral for x given by 
\begin{equation}\label{derive}
PI[x] = \int(dx)e^{\frac{i}{\hbar}S{_{eff}} [x]}
\end{equation}
where
\begin{equation}\label{expand}
S{_{eff}}[x]=\frac{1}{2}\int dt\left(m\frac{d^2x}{dt^2}+\eta\frac{dx}{dt}+kx+\frac{dV}{dx}\right)^2
- i tr\ell{n}\left(m\frac{d^2}{dt^2}+\eta\frac{d}{dt}+k+\frac{d^2V}{dx^2}\right).
\end{equation}

A perturbation expansion using the Green's functions $g(t-t')$ and $h(t-t')$ can be derived from equations (\ref{derive}) and (\ref{expand}).

In conclusion, let us put in context what was accomplished in this paper.  To the extent that there exist classical, open, dissipative systems that are subject to a random force as given by equation (\ref{open}), this paper can be used to directly quantize such systems.  But equation (\ref{open}) describes a ``phenomenological'' system, which should be derivable from an underlying system and bath dynamics.

The underlying dynamics is a matter of building models appropriate to particular systems.  In general, different models will have different realizations of dissipative and random forces.  For example, Caldeira and Leggett \cite{cald} modelled the environment with a bath of different oscillators that have an unrealistic interaction with the particle (interaction does not fall-off with distance and not translation-invariant).  In spite of these shortcomings the random and dissipative forces result from a quantum-statistical treatment of the system-bath dynamics when the oscillators have a particular frequency distribution.

We now ask, should the ``phenomenological'' theory be quantized directly or should we start with the system + bath dynamics?  Is the quantum theory of the phenomenological dynamics the same as that of the quantum theory of the system + bath dynamics then integrating out the bath degrees of freedom?  In other words, are the two pathways (see figure) to the quantum theory of the system lead to the same result?  In the case of the Caldeira-Leggett model, the two pathways give the same result as both yield the same quantum Fokker-Planck equation.
\begin{figure}
\centering
\includegraphics[totalheight=4in]{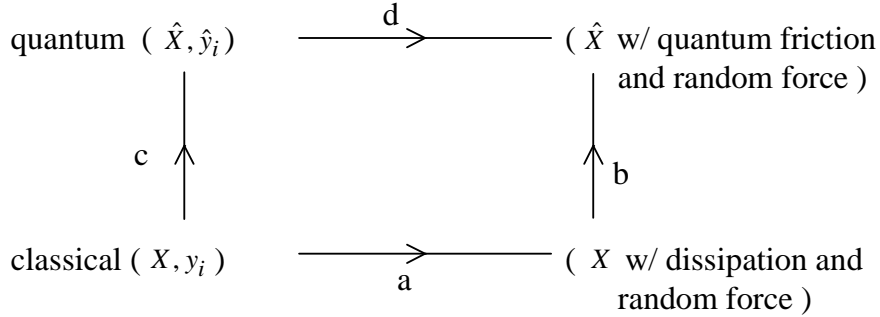}
\caption{Two pathways to quantum theory for an open, dissipative system subject to a random force (degree of freedom given by X).  The bath degrees of freedom are specified by $(y_i)$.  Starting from  the underlying classical dynamics for $(X,y_i)$, the two pathways are given by (1) - steps (a) and (b); and (2) - steps (c) and (d)}
\end{figure}
In the case of a slow degree of freedom interacting with a few, chaotic degrees of freedom the equivalence of the two pathways is not apparent.  Berry and Robbins \cite{robbins}, using an expansion based on the small parameter $\epsilon = \frac{\tau_f}{\tau_s}$ (ratio of the characteristic times of the slow and fast dynamics), showed that in the first-order there is classical dissipation but no quantum dissipation.  This apparently violates the correspondence principle and possibly is an example of a system where the two pathways do not yield the same result.  However, as we will show below, these observations are not correct.  As for the apparent violation of the correspondence rule, a careful accounting of the two small parameters $\hbar$ and $\epsilon$ by Wilkinson \cite{wilkin} showed that the classical dissipative force can be derived from the quantal Landau-Zener transitions.  This means there is no violation of the correspondence principle.

Similarly, it is not unreasonable to expect that the two pathways will yield the same quantum theory for the following reasons.  Generally, step (a) involves a statistical treatment using the Liouville distribution function.  Integrating out the bath degrees of freedom will give the reduced distribution function of X, which is used to compute the average of X, i.e., $\langle{X}\rangle$, and its dynamics.  This should result in a classical dynamics which is dissipating and subject to a random force.  Step (b), the direct quantization of an open, dissipative system, can be done either through the methods of Kostin, Yasue and this paper.

Step (c), on the other hand, involves the density matrix, which satisfies the quantum version of the Liouville equation.  Following this by step (d), which involves integrating out the bath degrees of freedom, results in the quantum theory of X.

Since the classical limit of step (c) is step (a), then as long as we account properly the bath degrees of freedom in steps (b) and (d), the two pathways should yield the same result.  The dominant configurations that give fluctuation and dissipation may not be the same for steps (b) and (d) as the Berry and Robbins example shows.  In this example, classical dissipation is due to the transfer of energy from the slow degree of freedom to all the fast, chaotic degrees of freedom.  In the quantum case, dissipation is a collective effect of avoided crossings described by the Landau-Zener theory.

To substantiate the above conjecture, let us make use of the Berry and Robbins' example.  Step (a) was carried out in the papers of Berry and Robbins and Jarzynski \cite {jarzynski}.  Although the full Liouville distribution was not explicitly worked out (only the zeroth order distribution was given by Berry and Robbins in terms of the microcanonical ensemble while Jarzynski used an arbitrary function of the interaction hamiltonian subject to a consistency condition) the effective forces (Born-Oppenheimber, geometric magnetism and dissipative) acting on the slow degree of freedom were derived.  Jarzynski's paper worked out the forces by deriving the Fokker-Planck equation satisfied by the reduced phase space distribution of the slow degrees of freedom.  

Steps (c) and (d), on the other hand, were worked out in detail by Berry and Robbins and by Wilkinson.  Berry and Robbins' construction of the density matrix lends easy identification of the classical limit.  Although their expression for the quantal ($\hbar$ finite) dissipative force in order $\epsilon$ vanishes, Auslaender and Fishman \cite {auslaender} showed that the classical value is derived if the limit $\hbar\to 0$ is taken first before $t\to\infty$ is taken.  This step is equivalent to the reverse of step (b).

In essence, the cited works show that steps (c), (d) and reverse of (b) is the same as step (a).  It would be more desirable to show that following step (a) by step (b) gives the same quantum theory for the slow degree of freedom as steps (c) and (d). Since the effective dynamics of the slow degree of freedom does not involve a random force, the best way to carry out step (b) is by quantizing the classical Fokker-Planck dynamics derived by Jarzynski.  Full equivalence of steps (c) and (d) to steps (a) and (b) is established if the quantum theory is the same as the Wigner distribution derived from the reduced density matrix.

Finally, as a corollary to the above result, it must follow that if two baths characterized by the sets of degrees of freedom $\{y_i\}$ and $\{z_i\}$ yield exactly the same classical dynamics with dissipation and random force, then the quantum theory of the two system + bath dynamics followed by integrating out their respective bath degrees of freedom will also be the same.
														
Acknowledgement:

	This work was started when the author visited the Institut fur Physik of the University of Mainz.  The author is grateful to Martin Reuter for hosting his stay and the Alexander von Humboldt for financial support. The author would like to thank Pol Nazarea for bringing reference 7 to his attention.  This research is supported in part by the Natural Sciences Research Institute of the University of the Philippines.

\end{document}